\documentclass[reprint,amsmath,amssymb, aps]{revtex4-1}
\usepackage{natbib}
\usepackage{graphicx}
\usepackage{braket}
\usepackage{dcolumn}
\usepackage{bm}
\usepackage{hyperref}
\usepackage[font=small,labelfont=bf]{caption}
\DeclareCaptionFont{small}{\small}

\begin{document}

\title{Entanglement and Quantum phase transition in topological insulators }

\author{Prasanta K Panigrahi}
\email{panigrahi.iiser@gmail.com}
\affiliation{Indian Institute of Science Education and Research Kolkata, Mohanpur, 741246, West Bengal, India}

\author{Bhavesh  Chauhan}%
\affiliation{Physical Research Laboratory, Navrangpura, Ahmedabad, 380009, Gujarat, India.}

\author{Anvesh Raja K}
\affiliation{Indian Institute of Science Education and Research Kolkata, Mohanpur, 741246, West Bengal, India}%

\author{Anant Vijay}
\affiliation{Indian Institute of Science Education and Research Kolkata, Mohanpur, 741246, West Bengal, India}
\date{\today}

\begin{abstract}
Presence of entangled states is explicitly shown in Topological insulator (TI) $Bi_2Te_3$. The surface and bulk state are found to have the different structures of entanglement. The surface states live as maximally entangled states in the four-dimensional subspace of total Hilbert space (spin, orbital, space). However, bulk states are entangled in the whole Hilbert space. Bulk states are found to be entangled maximally by controlled injection of electrons with momentum only along the z-direction. Scheme to detect entanglement in a 2-D model using measurement, confirming natural implementation of universal Hadamard with Controlled-NOT gates is explicated.

\end{abstract}

\maketitle

\section{\label{sec:level1} Introduction}

Creating and entangling single qubits, their scalability and protection against decoherence are key to the realization of quantum devices and quantum computers. In this regard, superconducting qubits \cite{scq}, quantum dots \cite{qd} and nitrogen defect in diamond have shown promise \cite{qcr}. In the absence of perfect isolation from surroundings, the above systems are prone to decoherence, which limits their applicability \cite{tqc}. In recent times, topological quantum computation with the underlying states protected by topology has attracted attention because of its robustness against decoherence \cite{qz}. Interestingly, topological insulators exhibit topologically protected surface states \cite{Zhang,Mele,Qi,Fu,qz}, and have found applications in spintronics \cite{sds1} and electrical memory devices \cite{sds, mem1, mem2}. Here, we demonstrate realization of entangled qubits and controlled variation of entanglement with parameter tuning. For specificity we have considered $Bi_2Te_3$, however, our approach is applicable to other 3-D gapped topological insulators. \vspace{.2cm}

Topological insulators are characterized by wave-functions with coupled spin, orbital, and spatial degrees of freedom. Entanglement between orbital and spin degree of freedom naturally arises in such systems due to spin-orbit coupling. Consequently, level crossing occurs between corresponding pairs of states. A quantum phase transition (QPT) separates the topologically non-trivial phase, from its trivial counterpart. The nature of coupling of the three degrees of freedom is expected to be different for the conducting surface and the insulating bulk state, and also in trivial and non-trivial phases. In case of $Bi_2Te_3$, one can project the system into a subspace spanned by the the four states $\Ket{P1^+_{-}, + \frac{1}{2}}, \Ket{P2^-_{+}, + \frac{1}{2}},  \Ket{P1^+_{-}, - \frac{1}{2}},\Ket{P2^-_{+}, -\frac{1}{2}}$ with the kinetic term (spatial part) arising perturbatively through the $\vec{k}\cdot\vec{p}$ perturbation expansion \cite{zhang1}. This results in Dirac type Hamiltonian with a Clifford algebra structure. Keeping in mind the entangled structure of the Hilbert space for TI and its role in QPT, we carry out a systematic investigation of the parameters affecting the entanglement and its behavior in trivial and non-trivial phases. It is also required for their possible use in quantum computation and other device applications. Here, we explicate the formation of entangled states in the 3D TI $Bi_2Te_3$  model.

   The paper is organized as follows: In Sec-II we present the model for $Bi_2Te_3$ and obtain energy spectra for surface and bulk states. Sec-III deals with QPT at $\Gamma$ point and entanglement characteristics as a function of Hamiltonian parameters. Sec-IV explicates a scheme to study entanglement using conductance measurement in a  2-D system of the underlying state. The last section is devoted to concluding remarks and future directions of work.

.

\section{\label{sec:level3} Model Hamiltonian}

The minimum model Hamiltonian for $Bi_{2}Te_{3}$ with four states $\Ket{P1^+_{-}, + \frac{1}{2}}, \Ket{P2^-_{+}, + \frac{1}{2}},  \Ket{P1^+_{-}, - \frac{1}{2}},\Ket{P2^-_{+}, -\frac{1}{2}}$ as basis, can be written as \cite{bhz,Dai, ti}: 

\begin{equation}\label{eqn:KM}
 H({\bf k}) = \epsilon({\bf k}) + \begin{bmatrix}
M({\bf k}) & A_2k_+ & 0 & A_1k_z\\ 
A_2k_- & - M({\bf k}) & A_1k_z & 0 \\ 
0 & A_1k_z & M({\bf k}) &-A_2k_-\\ 
A_1k_z & 0 & -A_2k_+  &-M({\bf k})
\end{bmatrix}
\end{equation}

where $k_\pm = k_+ \pm ik_{y}$ and, 
\begin{eqnarray}
\epsilon({\bf k}) &= C + D_1({k_{z}^{2}}) + D_2({k_{z}^{2}} + {k_{y}^{2}}) \\
M({\bf k})&= M-B_1({k_{z}^{2}}) - B_2({k_{z}^{2}} + {k_{y}^{2}}) \\
\end{eqnarray} 

It differs from the Dirac Hamiltonian as it contains the parabolic band term $Bk^{2}$, changing the $Z_{2}$ topological index from zero to one \cite{sqshen}.
Here, $ P1_-^+$ and $P2_+^-$ are two hybrid orbitals near the Fermi surface.
\begin{figure}[h!]
    \includegraphics[width=6cm]{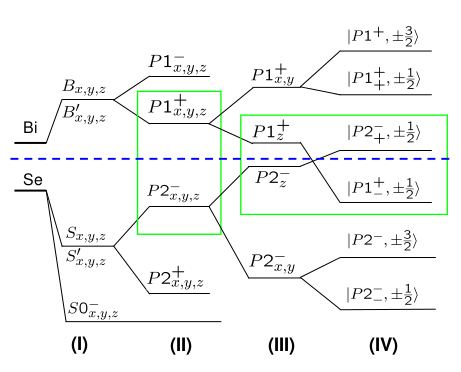}
    \caption{Schematic diagram of the band structure of $Bi_{2}Se_{3}$, depicting four steps of energy levels splitting: (I) the hybridization  (II) bonding and anti-bonding (inversion symmetry), (III)  crystal field splitting and (IV) the SOC.}
    \label{BI}
\end{figure}
Due to larger principal quantum number of Bi compared to Te, its energy levels lies in conduction band. The total angular momentum along the z-direction is conserved after taking into account spin-orbit coupling (SOC). Hybridization only occurs between the states $\Ket{\Lambda,p_{z},\uparrow} \& \Ket{\Lambda,p_{+},\downarrow}$ and $\Ket{\Lambda,p_{z},\downarrow} \& \Ket{\Lambda,p_{-} ,\uparrow}$ (where $\Lambda= P1_-^+, P2_+^- $), which leads to level crossing between pair of states $\Ket{P1_{-}^+, \pm{} \frac{1}{2}}$ and $\ket{P2_+^-,\pm{}\frac{1}{2} }$ (see FIG.\ref{BI}).

For the bulk eigenstates, we start with the ansatz $\Psi(k)= e^{-i(\omega t -k_{x}.x-k_{y}.y)} \phi_j(k_{z})$, which comprises of plane waves along x- and y- direction and four component spinorial part $\phi_j(k_{z})$. Using the fact that the electrons can be injected in the media in a particular momentum state such that the system is in the $k_{z}$ eigenstate ($k_{x}=k_{y}=0$), energy dispersion for \eqref{eqn:KM} is given by $E = \pm{}\left[ (M-Bk_{z}^{2})^{2}+ (Ak_{z})^{2} \right]^{1/2}$. Two orthogonal doubly degenerate eigenstates $\phi_j(k_{z})$ for this system corresponding to eigenvalues  $\pm E$ are given as:\\

$\begin{pmatrix}
        \frac{M({\bf k_{z}}) \pm  E}{Ak_{z}} & 0 & 0 & 1 
\end{pmatrix}^{T}$;
$\begin{pmatrix}
 0 & \frac{M({\bf k_{z}}) \pm  E}{Ak_{z}}  & 1 & 0
 \end{pmatrix}^{T}$. 
\\
\begin{figure}
    \includegraphics[width=4.0cm]{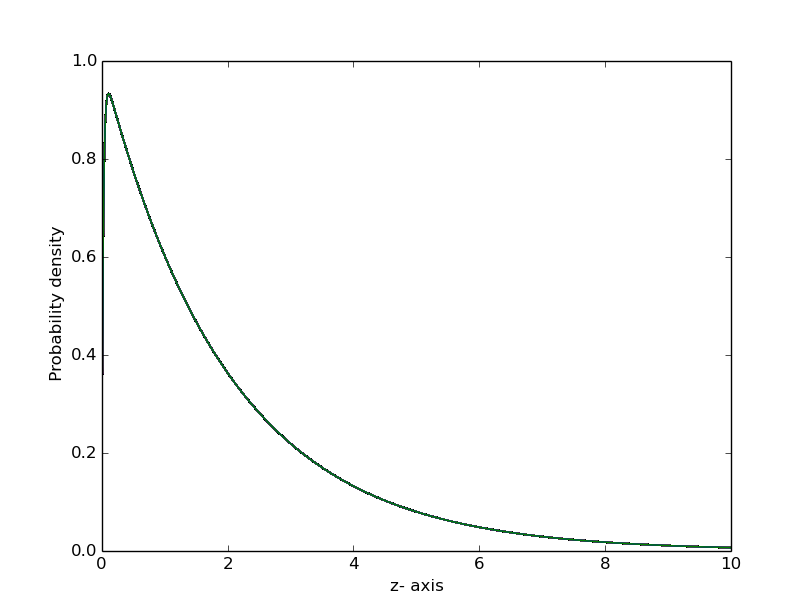}
    \captionsetup{font=small}
    \caption{\textit{Probability distribution of the surface state along z-axis for parameters values $A_{2}= 4, M= 2 $\&$ B= 0.1$ .}}
\end{figure}\\
These states are not separable in any of the three different subspaces. The surface states (zero energy) solutions can be obtained by applying boundary condition in real space coordinates normalized to half surface $0\leq z \leq \infty$: \\
$\Psi(x,y,z,t)= N_{s}\phi_{1,2}(z) e^{-i (p_{x}x - p_{y}y )}$,\\
  where $\phi_{1}(z)$  = $(e^{-\lambda_{-}z}-e^{-\lambda_{+}z})$ 
  $\begin{pmatrix}
 0 & \pm i  & 1 & 0
 \end{pmatrix}^{T}$;\\
 $\phi_{2}(z)$  = $(e^{-\lambda_{-}z}-e^{-\lambda_{+}z})$ 
  $\begin{pmatrix}
 1&  0  & 0 &  \pm i
 \end{pmatrix}^{T}$ and\\ $N_{s}= \frac{\sqrt{\lambda_{+}\lambda_{-}(\lambda_{+} + \lambda_{-})}}{(\lambda_{+} - \lambda_{-})}$; $\lambda_{\pm{}} = \frac{A}{2B} \pm{} \frac{\sqrt{A^{2}-4MB}}{2B}$. Curvature parameter $B$ controls the location of this zero energy state from the boundary. Increasing $B$ from 0.1 to 1 shifts $\left | \Psi \right |_{max}$ from 0.1 to 0.6 along +z-axis and $\left | \Psi \right |_{max}$ value decreases to half, as depicted in FIG.2.

\section{\label{sec:level2} Quantum Phase Transition (QPT) and Entanglement}

We now analyse entanglement properties of finite energy bulk states, as the zero energy states are maximally entangled in the four dimensional subspace, being separable in spatial degree of freedom. As is well known entanglement has close  connection  with quantum phase transition (QPT) {\cite{22,26}}, which  occurs when the ground state of a system changes by varying parameters such as magnetic field, pressure, etc {\cite{S,R}}.  
This leads to change in the symmetry of the ground state. For the above mentioned model symmetry changes by band closing and reopening as sign of $M/B$ is changed (see FIG.3. below). 

\begin{figure}[h!]
    \includegraphics[width=2.5cm]{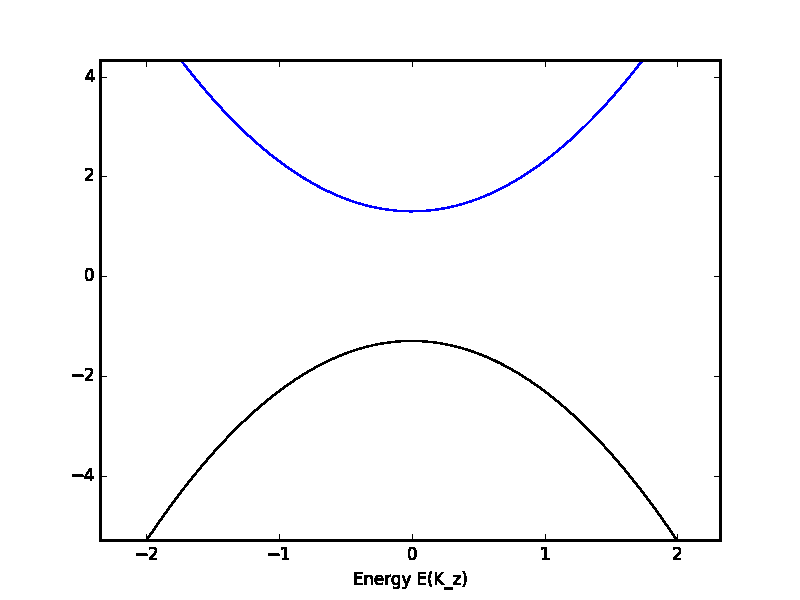}
    \includegraphics[width=2.5cm]{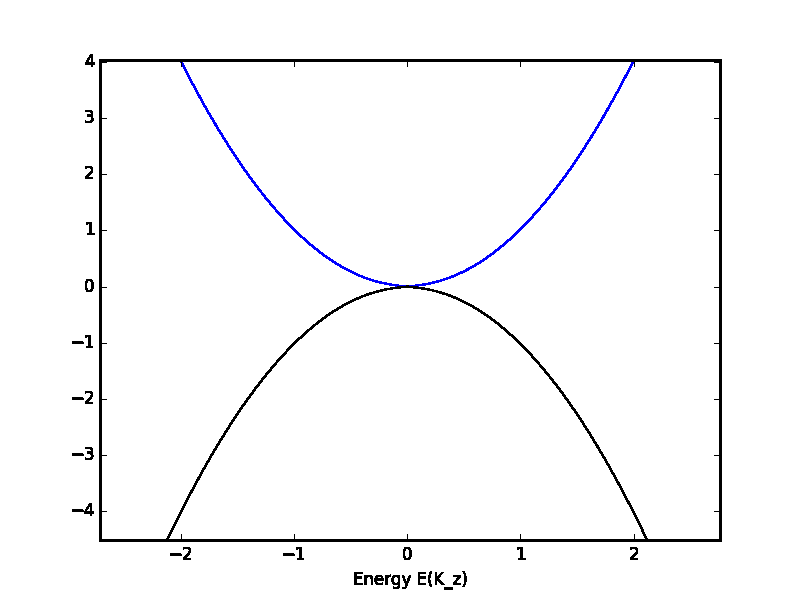}
    \includegraphics[width=2.5cm]{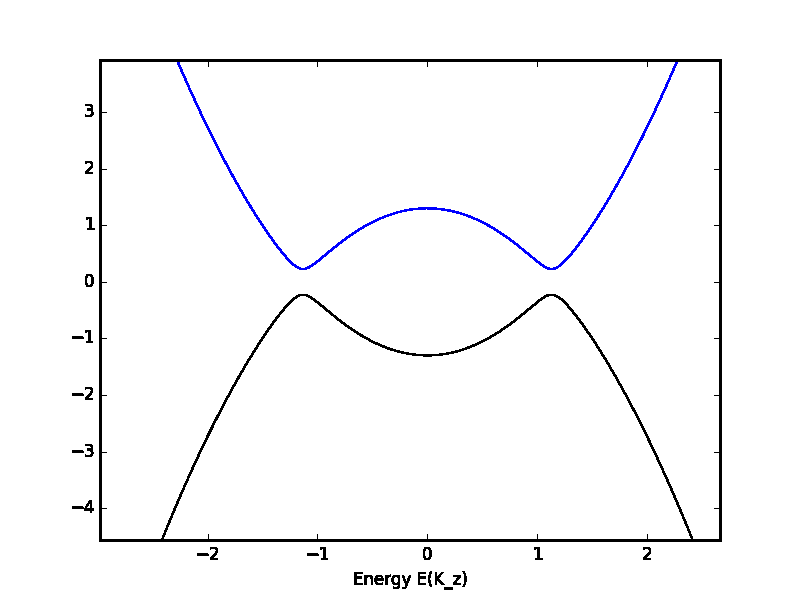}
    \captionsetup{font=small}
    \caption{\textit{Plot showing the band structure  along z-direction. Gapped band structure before QPT i.e., $M/B<0$ (left), at QPT $M/B=0$ (center) and band inversion for case $M/B >0$ ; with $M= \pm1.3  \&  0; B=1; A=.2$.} }
\end{figure}
\subsection{Concurrence and quantum phase transition}
Among the several measures to extract the signature of QPT \cite{Fu, Fu1,hk} in solid state systems mentioned in \cite{sri,pkp2}, we employ concurrence \cite{woot} as tool \cite{chiru3}. It gives the amount of state overlap and for the present system is given by (for both $E,-E$):

\begin{equation}
\mathcal{C}_{\pm}= \frac{1}{2}max\{0, \frac{1}{\sqrt{N_{\pm}}}(\frac{M({ \bf k})\pm E}{Ak})\} 
\end{equation}   

The orthogonal states mentioned in last section  are pure states and can be designated as follows:

\begin{eqnarray}
\phi_{\pm 1}(k)=\frac{1}{\sqrt{N_{\pm}}}(a_{\pm}\Ket{00}+ \Ket{11})=\frac{1}{\sqrt{N_{\pm}}}\begin{pmatrix}
a_{\pm} \\ 
0\\ 
0\\ 
1
\end{pmatrix} \\\phi_{\pm2}(k)= \frac{1}{\sqrt{N_\pm}}(a_{\pm}\Ket{10}+\Ket{01}) = \frac{1}{\sqrt{N_\pm}}\begin{pmatrix}
0 \\ 
a_{\pm} \\ 
1\\ 
0
\end{pmatrix}
\end{eqnarray} 

With $N_\pm= 1 + a_{\pm}^2$ and $a_{\pm}=\frac{M({\bf k}) \pm E}{Ak} $ being normalization constants. One can see in the concurrence plot (FIG.4. upper panel) that for small values of B, it increases and attains the maximum value of one at the critical point $B_{c}$. $B_{c}$ corresponds to phase transition point. For higher values all states become separable, thus changing the ground state of the system and revealing the presence of QPT. Changing the sign of M (black curve) is equivalent to phase transition.

\begin{figure}[h!]
    \includegraphics[width=5cm]{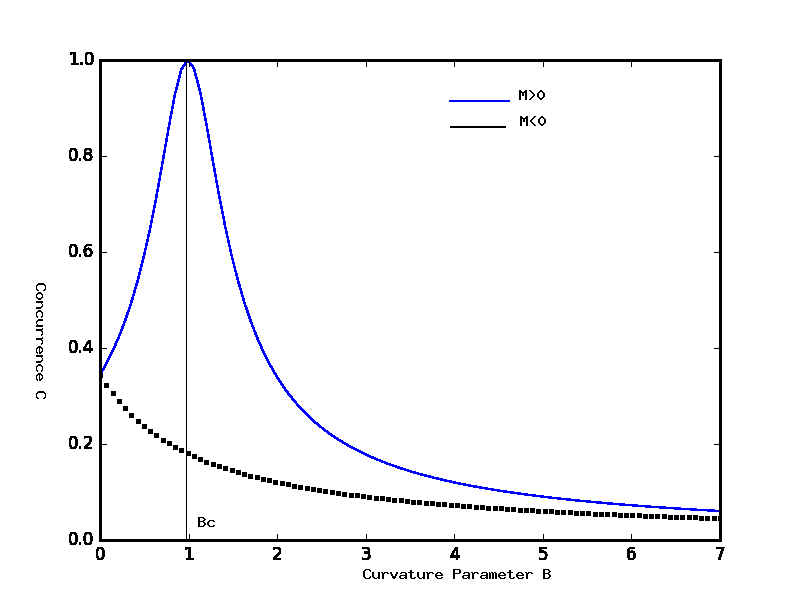}
    \includegraphics[width=5cm]{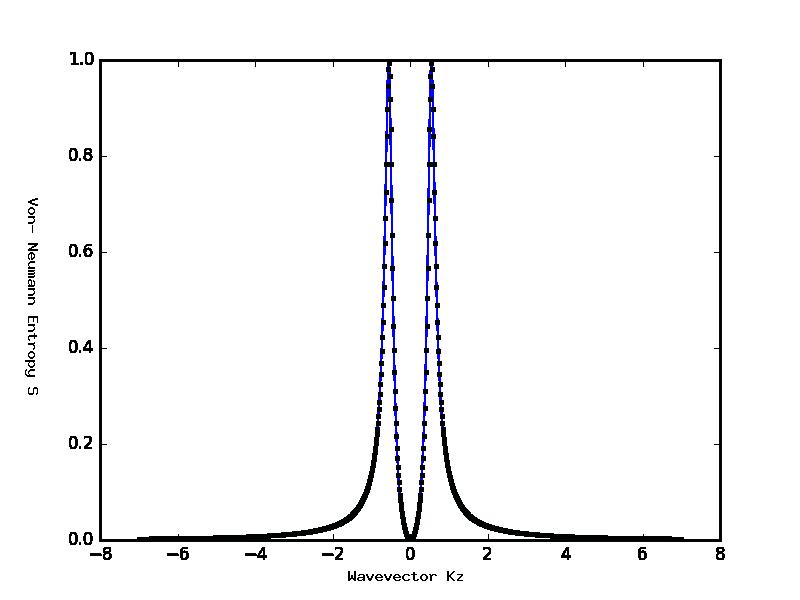}
   
    \captionsetup{font=small}  
   \caption{\textit{ Concurrence plot as a function of $B$ (upper panel). Dark line corresponds to $M>0$ and dashed line corresponds to $M<0$ respectively(above plot). Entropy variation with wave vector $k_{z}$ showing maxima at $\pm\sqrt(\frac{M}{B})$ (below plot).}}
   \label{f2}
\end{figure}

In summary, $M$ and $B$ are the parameters that controls phase transition and concurrence ($\mathcal{C}$) is maximum if $M/B>0$. Parameter $M$ represents the mass term, which can be tuned by external electric field or doping, whereas $B$ is the curvature parameter. This Hamiltonian describes a trivial insulator for $\frac{M}{B} < 0$. However, when $\frac{M}{B} > 0$ the bands are inverted leading to a TI.
It may be noted that von Neumann entropy $\rho  =(\frac{1}{\sqrt{N}})^{2}\log_2(\frac{1}{\sqrt{N}})^{2} $ provides the same results and we get entropy maxima at momenta $k = \pm{\sqrt{\frac{M}{B}}}$. The corresponding states at these values become $\phi_{1}(z)$  = $\sqrt{\frac{1}{2}}$
  $\begin{pmatrix}
 0 & \pm 1  & 1 & 0
 \end{pmatrix}^{T}$ (Bell states).
.  

\section{\label{sec:level4}Gate Implementation}

Utilization of the topological states for practical purposes requires measurement, which confirms the existence and amount of entanglement. We consider a 2-D TI model to describe a measurement scheme (see FIG.6.). The effective Hamiltonian for the model can be written as $\begin{bmatrix}
h_{+} & 0\\ 
0 & h_{-}
\end{bmatrix}$, with $h_{\pm} = vp_{x}\sigma_{x} \pm vp_{y}\sigma_{y} + (mv^2-Bk^2)\sigma_{z}$.\vspace{1mm} At small values of $k_{x}$, zero energy solutions take the form  $\Psi(x,y,z,t)_{\pm}= N_{s}\phi(y) e^{-ik_{x}x }$ \, \& $\phi(y)$ = $(e^{-\lambda_{-}y}-e^{-\lambda_{+}y})$   $\begin{pmatrix}
 0 & \pm i  & 1 & 0
 \end{pmatrix}^{T}$.\vspace{1mm} Here, $\sigma_{i}$'s do not represent real spin. However, $\Psi(x,y,z,t)_{\pm}$ are almost polarized along one direction of electron spin. Hence, Pauli matrices can be regarded approximately as real spin matrices. These edge states then can be distinguished as orthogonal eigenstates of helicity operator  $\Sigma = \tau_{y}\otimes \sigma_{x}$ 
 \cite{Germany}. \\

 $ \Sigma \Psi(x,y,z,t) = \tau \Psi(x,y,z,t)$   \qquad  $\tau = \pm{1}$.

\begin{figure}
   \includegraphics[width=4cm]{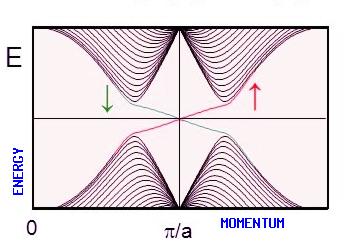}
\captionsetup{font=small}    
    \caption{\textit{Schematic of band dispersion in 2-D lattice. Two crossing (red \& green) branches correspond to a pair of edge states with opposite spin helicities $τ = \pm{1}$.}}
    \label{f3}
\end{figure}

\begin{figure}
   \includegraphics[width=4cm]{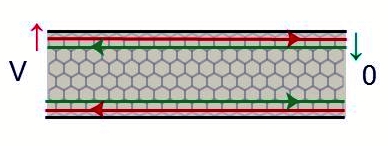}
\captionsetup{font=small}    
    \caption{\textit{Schematic diagram of 2-D TI ribbon showing the spin-up and spin-down states in terms of entangled states.}}
    \label{f4}
\end{figure}
In a small-scale semiconductor with a few modes, quasi-Fermi levels for $-k_{x}$ states and  $+k_{x}$ states are notably different \cite{SupDat}.  Net current flows along $+x$ direction due to difference between quasi-fermi levels of  $ \Ket{+k_{x},\uparrow}$ and $ \Ket{-k_{x},\downarrow} $ edge states, when potential V  is applied across the left side (FIG.6).  For measurement one can choose a spin filter (which is equivalent to choosing a basis) at the right end followed by a current measurement. Allowing down spin state through the spin filter would then result in zero current. On the contrary, choosing up spin state would result $\frac{e^2}{h}$ value \cite{konig1} as current measurement.  A large number of repetitive measurements can confirm the existence of maximally entangled states.

\section{\label{sec:level4}Conclusion}
In conclusion, the model Hamiltonian describing a 3-D topological insulator $Bi_{2}Te_{3}$ can host entangled states. 

Surface states are maximally entangled in a sub-space, while bulk states are entangled in the whole space.
However, we conclude that it is possible to realize Bell states in the bulk by controlled injection of electrons, at phase transition point. Measurement scheme shown using a 2-D model implies, natural implementation of quantum gates. Further investigations are required for non-destructive measurements of such states.

\bibliographystyle{unsrtnat}
\bibliography{bibfile}
 
\end{document}